\documentstyle[12pt,epsf]{article}
\newcommand{\gtap}{\;{\raise.3ex\hbox{$>$\kern-.75em\lower1ex\hbox{$\sim$}}}\;}
\newcommand{\ltap}{\;{\raise.3ex\hbox{$<$\kern-.75em\lower1ex\hbox{$\sim$}}}\;}

\begin{document}
\begin{titlepage}

\rightline{hep-ph/0201255}
\smallskip
\rightline{January 29, 2002}
\bigskip\bigskip
\begin{center}
{\Large \bf Dynamical symmetry breaking of $\lambda\phi^4$ theory
in two loop effective potential} \\
\medskip
\bigskip\bigskip\bigskip\bigskip\bigskip\bigskip
{{\bf Ji-Feng Yang} and {\bf Jian-Hong Ruan}}\\
\medskip
\bigskip\bigskip
\em{Department of Physics, East China Normal University, Shanghai
200062, China} \\
\bigskip
\end{center}
\bigskip\bigskip\bigskip

\begin{abstract}
The two loop effective potential of massless $\lambda\phi^4$
theory is presented in several regularization and renormalization
prescriptions and the dynamical symmetry breaking solution is
obtained in the strong-coupling situation in several prescriptions
except the Coleman-Weinberg prescription. The beta function in the
broken phase becomes negative and the UV fixed point turns out to
be a strong-coupling one, and its numeric value varies with the
renormalization prescriptions, a detail which is different from
the asymptotic free solution in the one loop case. The symmetry
breaking phase is shown to be an entirely strong-coupling phase.
The reason of the relevance of the renormalization prescriptions
is shown to be due to the nonperturbative nature of the effective
potential. We also reanalyzed the two loop effective potential by
adopting a differential equation approach based on the
understanding that takes all the QFT's as the ill-defined
formulations of the 'low energy' effective theories of a complete
underlying theory. Then the relevance of the prescriptions of
fixing the local ambiguities to physical properties like symmetry
breaking is further emphasized. We also tentatively proposed a
rescaling insensitivity argument for fixing the quadratic
ambiguities. Some detailed properties of the strongly coupled
broken phase and related issues were discussed.
\end{abstract}
\end{titlepage}

\section{Introduction}
The Standard model (SM) has now been firmly established with most
of its predictions experimentally confirmed. New physics beyond
the SM are being intensively explored from theoretical
perspective, but no concrete experimental evidences has yet been
found. A major motivation to go beyond the SM has been to get rid
of those theoretically unsatisfactory aspects of the SM such as
the hierarchy or naturalness problem\cite{Hie} and the
triviality\cite{Trivial} of the Higgs sector, and that there are
too many parameters to be explained. Thus most particle theorists
believe that the SM is only an effective theory of a fundamental
theory. The currently prevailing direction to go beyond the SM has
been the string theory\cite{String} and/or supersymmetric field
theories\cite{SUSY}. These theories modify the SM profoundly. As a
matter of fact, the most demanding task in and beyond SM physics
is to find the true mechanism of symmetry breaking to replace the
Higgs sector that suffers the above-mentioned defects and is held
as phenomenological. In this connection, there has been another
important theoretical direction that does not modify the SM so
profoundly, the technicolor model and its descendants\cite{TC}.
All the above theoretical constructions share a common feature:
the elementary Higgs scalar fields are excluded and the solution
to the hierarchy and triviality problem must be in
non-perturbative regime\cite{SUSY}.

However, more than a decade ago, there were some efforts to revive
the $\lambda\phi^4$ interaction from the perturbative triviality
by showing that the one loop effective potential of the massless
$\lambda\phi^4$ permitted an nontrivial non-perturbative
renormalization\cite{Stevenson}, i.e., $\beta(\lambda)<0$, in
contrast to the perturbative renormalization where
$\beta(\lambda)>0$ (leading to triviality). On the other hand, it
has been recently proposed that color confinement is closely
related to flavor symmetry breaking\cite{Locking} and even that
the color symmetry be realized via Higgs
mechanism\cite{Wetterich}. In a sense, the Higgs model or the
$\lambda\phi^4$ interaction is still useful and should be further
explored to search for nontrivial solution of the model. If the
symmetry breaking can be dynamically realized together with
asymptotic freedom or non-triviality, then it will shed new light
on the confinement of color and symmetry breaking of the standard
model. Thus it is worthwhile to see if the interesting nontrivial
one loop solution can still exist after including higher loop
corrections or how it 'evolves' in presence of higher order
quantum corrections.

In this paper we provide a detailed report of our recent
investigation of the existence and new features (if any) of the
nontrivial dynamical symmetry breaking solution of the quartic
interaction by studying the two loop effective
potential\cite{DSB}. For convenience, we will consider the
simplest scalar model--the massless $\lambda\phi^4$ model with
$Z_2$ symmetry--with which the first example of the dynamical
symmetry breaking was demonstrated\cite{CW}. There is also a
technical concern in choosing massless scalar theory: there is a
nonconvexity in the tree interactions that affects the Higgs model
and often complicates the use of effective potential
methods\cite{Convex}, while in massless models the tachyon mass
term is gone and the configuration of the expectation value of
scalar field can be naturally interpreted as the homogeneous
argument of the effective potential.

In the meantime, we need to consider the regularization and
renormalization problems in the nonperturbative regime. Since the
effective potential is nonperturbative in nature, its
regularization and renormalization might become more subtle. There
have long been standard procedures to carry out perturbative
renormalization, but in nonperturbative contexts the
renormalization often needs to be dealt with case by case, example
by example. Moreover, the nonperturbative context sometimes allows
for an alternative renormalization solution, for example, the
nontrivial or asymptotic free solution for the one loop potential
of $\lambda\phi^4$ mentioned above\cite{Stevenson,Tarrach}. Other
examples about the subtleties associated with regularization and
renormalization can be found in the recent applications of the
effective-field-theory method\cite{EFT} to nucleon
interactions\cite{Kaplan,Maryland,Steele,Richardson}, where the
framework in use is necessarily nonperturbative. We hope our
experiences here might be useful in carrying out renormalization
within non-perturbative contexts.

The paper is organized in the following way. The two loop
effective potential will be given in dimensional and cutoff
regularization respectively in Sec.II. The bare and renormalized
effective potentials obtained in different schemes will also be
listed there. Then in Sec. III we investigate the existence and
the properties of the dynamical symmetry breaking solution(s) via
the effective potentials obtained various intermediate
renormalization prescriptions. There the prescription dependence
of the solution is exhibited and explained. Sec. IV will be
devoted to a new approach for evaluating the loop diagrams and the
relevance of the intermediate renormalization is highlighted. Some
properties and features of the symmetry breaking solution in the
two effective potential are also presented. Some discussions and
the summary will be given in the final section.

\section{Regularization and renormalization}
As is stated in the introduction, we will consider the massless
$\lambda\phi^4$ model with $Z_2$ symmetry: invariance under the
transformation of $\phi\rightarrow -\phi$. The algorithm for two
loop effective potential is well known according to
Jackiw\cite{Jackiw}
\begin{eqnarray}
\textsc{L}&=&\frac{1}{2}(\partial\phi)^2-\lambda\phi^4,\\
V_{(2l)}&\equiv& \lambda\phi^4+\frac{1}{2}I_{0}(\Omega)+3\lambda
I^2_1(\Omega)-48\lambda^2 \phi^2I_2(\Omega),\\
\Omega&\equiv& \sqrt{12\lambda\phi^2};\\
I_{0}(\Omega)&=&\int\frac{d^4
k}{(2\pi)^4}\ln(1+\frac{\Omega^2}{k^2});\\
I_1(\Omega)&=&\int\frac{d^4 k}{(2\pi)^4}\frac{1}{k^2+\Omega^2};\\
I_2(\Omega)&=&\int\frac{d^4 k d^4
l}{(2\pi)^8}\frac{1}{(k^2+\Omega^2)(l^2+\Omega^2)((k+l)^2
+\Omega^2)};
\end{eqnarray} Here we have Wick-rotated all the loop integrals
into Euclidean space. Let us calculate the three integrals in two
regularization schemes, dimensional and cutoff. As these integrals
have already been calculated in literature both in dimensional
regularization and in cut-off schemes, we will only need to list
the results here.

\subsection{Dimensional and cutoff regularizations}
In dimensional regularization, these integrals have been
calculated in the literature, see \cite{FJ}. Here we list the two
loop diagram (the sunset diagram) for example, the other integrals
will be delegated to Appendix A.
\begin{eqnarray}
\mu^{4\epsilon}I^{(D)}_2(\Omega)&=&\int\frac{\mu^{4\epsilon}d^D k
d^Dl}{(2\pi)^{2D}}
\frac{1}{(k^2+\Omega^2)(l^2+\Omega^2)((k+l)^2+\Omega^2)}\nonumber
\\
&=&-\frac{3\Omega^2}{2(4\pi)^4\epsilon^2}\{1+(3-2\overline{L})
\epsilon+ (2\overline{L}^2-6\overline{L}+7+6S-\frac{5}{3}\zeta(2)
)\epsilon^2\}
\end{eqnarray}with $S=\sum^{\infty}_{n=0}\frac{1}{(2+3n)^2},
\overline{L}=L+\gamma-\ln 4\pi$ and $L=\ln
\frac{\Omega^2}{\mu^2}$.

Similarly, in cutoff regularization, we find from
Ref.\cite{Jackiw},
\begin{eqnarray}
I^{(\Lambda)}_{2}(\Omega)&=&\int_{\Lambda} \frac{d^4k d^4l}
{(2\pi)^8}\frac{1}{(k^2+\Omega^2)(l^2+\Omega^2)((k+l)^2+
\Omega^2)} \nonumber
\\&=&\frac{1}{(4\pi)^4}\{2\Lambda^2-\frac{3\Omega^2}{2}\ln^2
\frac{\Omega^2}{\Lambda^2}+3\Omega^2\ln
\frac{\Omega^2}{\Lambda^2}+o(\Lambda^{-2})\}.
\end{eqnarray}
Note that the $\sim \Lambda^2$ term in the two loop integral is
not explicitly given in \cite{Jackiw}.

Note that the leading 'low energy' content of the sunset diagram
(the double log term) obtained in dimensional regularization
differs from that obtained in cutoff scheme. But this does not
matter, after subtracting the sub-divergences in such
diagrams\cite{FJ}, the 'nonlocal' term will be the
same\footnote{In a diagrammatic or perturbative framework, such
independence of regularization schemes is beyond doubt. But in a
nonperturbative framework, such independence is
controversial\cite{Kaplan,Maryland,Steele,Richardson,PH}. Since
our calculation is a systematic summation of infinite diagrams
(hence non-perturbative) where only a few diagrams are UV
ill-defined, this subtle point does not concern us here. There are
also some references\cite{Ordonez} where related issues are
discussed.}.

\subsection{Bare and renormalized effective potential}
With the preceding preparations, we can write down the bare
effective potential obtained, respectively, in dimensional and
cutoff regularizations:
\begin{eqnarray}
 V^{(D) }_{(2l)}( \Omega ) &=&\Omega ^{4}\{
\frac{1}{144\lambda }+\frac{-\frac{1}{\epsilon }
+\overline{L}-\frac{3}{2}}{ ( 8\pi ) ^{2}}+\frac{3\lambda }{( 4\pi
) ^{4}}[ (-\frac{1}{\epsilon }+\overline{L}-1) ^{2}+(
\overline{L}-1)^{2}\nonumber \\
&&+2( -\frac{1}{\epsilon }+\overline{L}-\frac{3}{2}) ^{2}+2(
\overline{L}-\frac{3}{2}) ^{2}+7+6S-\frac{5}{3}\zeta(2)] \};\\
V^{(\Lambda )}_{(2l)}(\Omega )&=&\Omega ^{4}\{ \frac{1}{144\lambda
}+\frac{ L_{\Lambda }^{\Omega }-\frac{1}{2}}{( 8\pi )
^{2}}+\frac{3\lambda }{( 4\pi ) ^{4}}[ (L_{\Lambda }^{\Omega
})^{2}-2+2( L_{\Lambda
}^{\Omega }-1) ^{2}] \} \nonumber \\
&&+\Omega ^{2}\{ \frac{2\Lambda ^{2}}{(8\pi )^{2}}+\frac{3\lambda
\Lambda ^{2}L_{\Lambda }^{\Omega }}{(4\pi )^{4}}-\frac{8\lambda
\Lambda ^{2}}{(4\pi )^{4}}\},
\end{eqnarray}where $L^{\Omega}_{\Lambda}=\ln
\frac{\Omega^2}{\Lambda^2}$. Here we have omitted all the field
independent terms. In the remaining part of this section, we focus
on the renormalization of $V^{(D) }_{(2l)}( \Omega )$ and
$V^{(\Lambda )}_{(2l)}(\Omega )$.

The renormalization will be done in $\overline{MS}$ scheme for
$V^{(D) }_{(2l)}( \Omega )$ (Cf.\cite{FJ}), while for $V^{(\Lambda
)}_{(2l)}(\Omega )$ the renormalization will be done in three
prescriptions: the one defined by Jackiw\cite{Jackiw}, the one
adopted by Coleman and Weinberg\cite{CW} and a new prescription,
$\mu^2_{\Lambda}$(a simulation of $\overline{MS}$, see Appendix
B). The results read
\begin{eqnarray}
V^{(\overline{MS}) }_{(2l)}( \Omega )
&=&\Omega^{4}\{\frac{1}{144\lambda}+\frac{\overline{L}
-\frac{3}{2}}{(8\pi)^2}+\frac{3\lambda}{(4\pi)^4}[3\overline{L}^2
-10\overline{L}+11+12S-\frac{8}{9}\pi^2]\};\\
V^{(\mu^2_{\Lambda})}_{(2l)}(\Omega )
&=&\Omega^{4}\{\frac{1}{144\lambda}+\frac{\tilde{L}
-\frac{1}{2}}{(8\pi)^2}+\frac{3\lambda}{(4\pi)^4}
[3\tilde{ L}^2-4\tilde{L}]\};\\
V^{(Jackiw)}_{(2l)}(\Omega )
&=&\Omega^{4}\{\frac{1}{144\lambda}+\frac{\check{L}}{(8\pi)^2}+
\frac{3\lambda}{(4\pi)^4}[3\check{L}^2-\check{L}]\};\\
V^{(CW)}_{(2l)}(\Omega )
&=&\Omega^{4}\{\frac{1}{144\lambda}+\frac{\breve{L}}{(8\pi)^2}+
\frac{3\lambda}{(4\pi)^4}[3\breve{L}^2-\breve{L}+\frac{205}{12}]\}
\end{eqnarray}with the notations defined as $\tilde{L}=
\ln \frac{\Omega^2}{\mu^2_{\Lambda}},\check{L}=\ln
\frac{\Omega^2}{12\lambda\mu^2_{Jackiw}}$ and $\breve{L}=\ln
\frac{\Omega^2}{12\lambda\mu^2_{CW}}-\frac{25}{6}$. In all the
above formulas the scheme dependence of field strength and
coupling constant are understood. Note that the $\mu^2_{\Lambda}$,
Jackiw and Coleman-Weinberg prescriptions were applied to the same
bare effective potential, i.e., that calculated in the cut-off
scheme.

\subsection{Prescription dependence}
Upon appropriate rescaling of the subtraction scales, all versions
of the effective potential take the following form (we will drop
all the dressing symbols)
\begin{eqnarray}
\label{SCH} V_{(2l)}( \Omega )
=\Omega^{4}\{\frac{1}{144\lambda}+\frac{L-1/2}{(8\pi)^2}
+\frac{3\lambda}{(4\pi)^4}[ L^2+2(L-1)^2+\alpha]\}
\end{eqnarray}with $L \equiv \ln \frac{\Omega^2}{\mu^2}$. Now we see the
explicit dependence of the effective potential upon the
intermediate renormalization prescriptions expressed by $\alpha$,
which varies across schemes as exhibited in Table~\ref{table1}.

Here the scheme dependence (regularization and/or renormalization)
of the effective potential as a nonperturbative quantity (summing
over infinite many one- and two-loop one particle irreducible
diagrams) differs from that of the perturbative
framework\cite{scheme} that arises from the truncation of
perturbation series (a sum of finite number of connected
diagrams)\footnote{Rigorously speaking, this property has been
established only in mass independent subtraction schemes or in
massless theories or in high energy region where mass effects are
negligible. The nontrivial influence of renormalization
prescriptions in defining masses has been recently
emphasized\cite{Pole} in theories with unstable elementary
particles (like $W^{\pm},Z^0$ bosons in electro-weak theory).}.
The difference in $\alpha$ could not be removed through
redefinition of the coupling constant (and perhaps of field
strength) without changing the functional dependence upon the
field expectation value, $\phi$. This is a crucial difference. The
main obstacles here are (i) the presence of the double log
dependence on $\phi$ (in $(\ln \frac{12\lambda \phi^2}{\mu^2})^2$)
and ii) the nonperturbative feature of the effective potential,
i.e., the sum over infinite many diagrams.

If one redefines the coupling constant and expand the new coupling
constant in terms of the old one like in the perturbative case
($\lambda^\prime=\lambda+a\lambda^2+b\lambda^3+\cdots$), one could
at best arrive at the other schemes' results {\em plus} extra
higher order terms that take the form $\sim \lambda^n \phi^4 \ln
\frac{\phi^2}{\mu^2}, n\geq3$. The same is true for the
redefinition of $\phi$ or $\Omega$. Since the effective potential
is nonperturbative in terms of $\lambda$ and $\phi$ in nature, one
should not discard such kind of higher order terms due to
consistency due to their nontrivial dependence upon $\phi$ that
will affect the symmetry breaking status, unlike in the
perturbative case. Otherwise, as will be clear shortly, even if
one put the consistency aside and discard such terms, the symmetry
breaking behavior \emph{will be changed} due to such kind of
redefinition and approximation. Thus even with the intermediate
renormalization done in the standard way, the nonperturbative
results depend on the prescriptions quite nontrivially. To the
best of our knowledge, this new feature in the nonperturbative
framework has not been explicitly and particularly pointed out.

If there are no double log terms present in the effective
potential except the single log terms, then the constant terms can
be easily redefined away or absorbed into the single log terms
without leading to new extra functional dependence upon $\phi$
that can affect the symmetry breaking. In gauge theories, there
are only single log terms present in the sum of one particle
irreducible diagrams. While here we encounter the essential
presence of double log terms in the sum of one-particle
irreducible diagrams at two loop level (recall that the effective
potential is the generating functional of the one-particle
irreducible diagrams), it is not difficult to see that a still
higher power of log terms can generally show up in higher loop
one-particle irreducible diagrams.

\begin{table}[t]
\caption{Values of $\alpha$ in various schemes}
\begin{center}
\begin{tabular}{c|c} \hline \hline &\\[-.4cm]
 Scheme &  $\alpha$\\ \hline &\\[-.4cm]
$\overline{MS}$& $-2.6878$\\
$\mu^2_{\Lambda}$& $-2$\\
Jackiw & $-\frac{5}{4} $\\
Coleman-Weinberg & $16\frac{1}{3} $\\
\hline \hline
\end{tabular}
\end{center}
\label{table1}
\end{table}
\section{Effective potential and symmetry-breaking solution}
Now let us start to determine the minima of the two loop effective
potentials that are renormalized in the prescriptions specified in
last section. We will work with the general parametrization form
of Eq.~(\ref{SCH}). Our goal is to solve the first order equation
\begin{eqnarray}
\frac{d V_{(2l)}( \sqrt{12\lambda\phi^2} )}{d\phi}=0
\end{eqnarray}
which becomes the following equation upon substituting
Eq.~(\ref{SCH}) into it,
\begin{eqnarray}
\label{SCH2}
24\lambda\phi\Omega^2[\frac{2V_{(2l)}(\Omega^2)}{\Omega^4}
+\frac{1}{(8\pi)^2}+\frac{3\lambda}{(4\pi)^4}(6L-4)]=0.
\end{eqnarray}

An obvious solution is $\phi=0$ which is the symmetric solution in
perturbative (weak-coupling) regime, while the existence of the
nonzero expectation value solution is determined by the existence
of a real number solution of $L$ to the following algebraic
equation
\begin{eqnarray}
\label{SCH3}
3L^2+(\frac{4\pi^2}{3\lambda}-1)L+\alpha+\frac{16\pi^4}{27\lambda^2}=0.
\end{eqnarray} Here it is obvious that the existence of real number
solutions depends on both $\alpha$ and $\lambda$. Since $\alpha$
is renormalization prescription dependent, it is natural to expect
that the solution and its existence are also prescription
dependent. Since symmetry breaking is a physical phenomenon, one
usually anticipates that the occurrence of symmetry breaking
should be independent of a manipulation of infinities, that is,
independent of renormalization schemes. Here we see a counter
example. In this connection, we would like to mention other
nonperturbative examples discussed in Ref.\cite{NILOU}, where the
physical predictions depend on the renormalization (and
regularization) prescription in use. The reason is basically the
same as has been given in the preceding subsection.
\subsection{Determinants of symmetry-breaking solution}
Now let us examine the symmetry-breaking solution in more detail.
Since we must start from a stable micropotential, the coupling
$\lambda$ must be a positive real number. Now let us closely
examine Eq.~(\ref{SCH3}). For Eq.~(\ref{SCH3}) to possess a finite
real number solution, we must impose the following criterion in
terms of $\alpha$ and $\lambda$
\begin{eqnarray}
\label{Delta}
\Delta\equiv
(\frac{4\pi^2}{3\lambda}-1)^2-12(\alpha+\frac{16\pi^4}{27\lambda^2})=
\frac{1}{3}[4-36\alpha-(1+\frac{4\pi^2}{\lambda})^2] \geq 0.
\end{eqnarray}
This inequality is only valid for certain ranges of $\alpha$ and
$\lambda$,
\begin{eqnarray}
\label{alpha}
\alpha &<& \frac{1}{12},\\
\lambda&\geq&
\lambda_{cr}\equiv\frac{4\pi^2}{\sqrt{4-36\alpha}-1}.
\end{eqnarray}

\begin{figure}[t]
\begin{center}
\vspace*{0cm} \hspace*{0cm} \epsfxsize=10cm \epsfbox{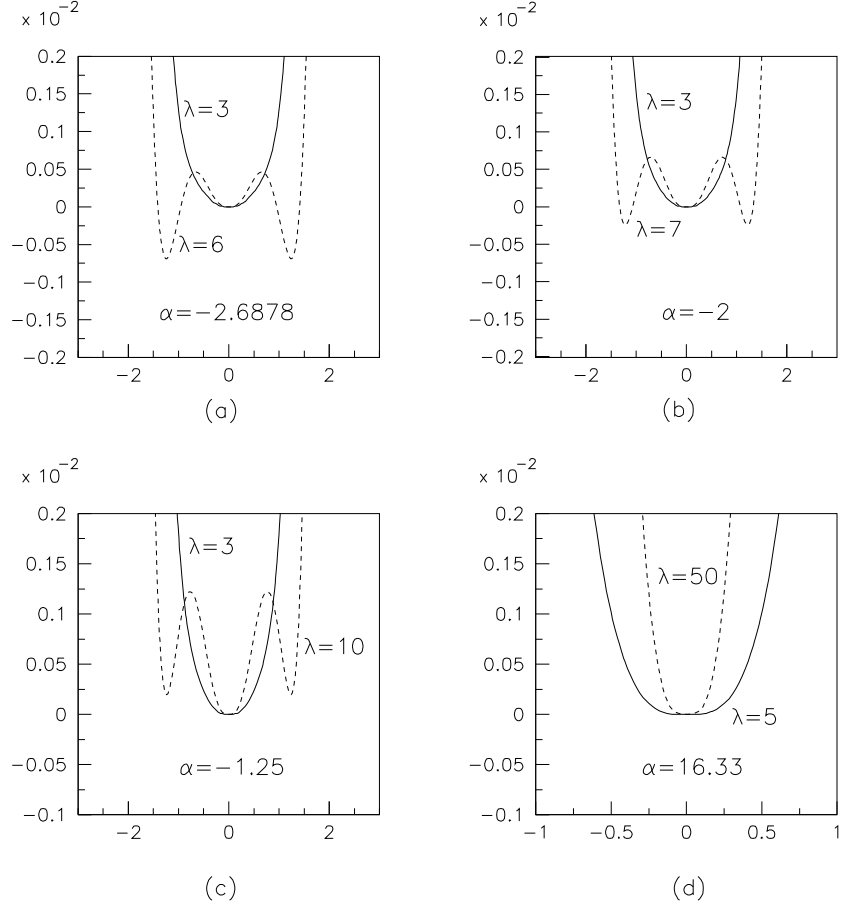}
\vspace*{-1.5cm} \caption{The two loop effective potential in
various renormalization prescriptions with different values of
coupling constant. In all the four prescriptions [(a)--(d)], the
horizontal axis represents the quantity
$\frac{\sqrt{12\lambda}\phi}{\mu}$ while the vertical axis
represents $\frac{V_{(2l)}}{\mu^4}$.} \label{V-1}
\end{center}
\end{figure}

Then the solutions to Eq.~(\ref{SCH3}) can be found provided the
above two requirements are satisfied in certain schemes,
\begin{eqnarray}
L_{\pm}(\lambda)=\frac{1}{6}[1-\frac{4\pi^2}{3\lambda}\pm
\sqrt{\Delta}].
\end{eqnarray}From this and the definitions $L \equiv \ln
\frac{\Omega^2}{\mu^2}, \Omega \equiv \sqrt{12\lambda\phi^2}$ we
can find the nonzero solutions of $\phi$, which read
\begin{eqnarray}
\phi^2_{\pm}(\lambda;[\mu,\alpha])=\frac{\mu^2}{12\lambda}
\exp\{\frac{1}{6}[1-\frac{4\pi^2}{3\lambda}\pm\sqrt{\Delta}]\}.
\end{eqnarray}
But the solutions corresponding to $L_{-}(\lambda)$ are local
maxima (tachyonic), only the $ L_{+}(\lambda)$ solutions are local
minima, this can be seen from the second-order derivative of the
effective potential at $\Omega^2_{\pm}$ (which is exactly the
effective mass),
\begin{eqnarray}
\label{mass} m_{eff;\pm}(\lambda)\equiv \frac{\partial^2
V_{(2l)}}{(\partial\phi)^2} \|_{\phi^2=\phi_{\pm}}=\pm
\frac{18\lambda^2\Omega^2_{\pm}}{(2\pi)^4} \sqrt{\Delta}.
\end{eqnarray}Because of the presence of the local maxima ($\pm
\sqrt{\frac{\mu^2}{12\lambda}}
\exp\{\frac{1}{12}[1-\frac{4\pi^2}{3\lambda}-\sqrt{\Delta}]\} $)
between the local minima $\phi=0$ and
$\pm\sqrt{\frac{\mu^2}{12\lambda}}
\exp\{\frac{1}{12}[1-\frac{4\pi^2}{3\lambda}+\sqrt{\Delta}]\}$,
the symmetry breaking must be a first-order phase transition when
it happens, in accordance with the recent results\cite{Agodi}
obtained through other approaches. It is also clear from
Fig.~\ref{V-1}, in which the shape of the effective potential is
depicted in several renormalization prescriptions ($\alpha$) for
different values of coupling constant.

The inequality~(\ref{alpha}) tells us that the renormalization
prescriptions do affect physical contents in the nonperturbative
framework: Not all prescriptions could be compatible with symmetry
breaking as far as the two loop effective potential is concerned
(the stability of such solutions will be discussed shortly). From
Table~\ref{table1} in Sec. II we see the following: For the two
loop effective potential, the Coleman-Weinberg scheme failed to
predict dynamical symmetry breaking as the critical
inequality~(\ref{alpha}) is badly violated there:
$\alpha_{CW}=16\frac{1}{3}=\frac{196}{12} \gg \frac{1}{12}$ while
the other three schemes do allow for symmetry breaking solutions.
The situation is not affected by the rescaling of the subtraction
points, one can check that even in the original form (Cf.Eq.(3.17)
in Ref.\cite{Jackiw}) the inequality corresponding
to~(\ref{Delta}) could not be satisfied (see Appendix C), in fact
the corresponding $\Delta$ is strictly negative for nonnegative
value of the renormalized coupling $\lambda$. The Fig.~\ref{V-1}
also exhibits such a prescription dependence.

Now we find a strong dependence of 'physical' properties upon
renormalization prescriptions, though it is demonstrated within a
model that is not quite realistic. It is not totally unexpected if
one recalls that the effective potential is a nonperturbative
object, as was noted in the preceding section. The only unexpected
point is that the pioneering prediction of dynamical symmetry
breaking has been done made in the Coleman-Weinberg scheme used in
the \emph{one loop} effective potential, while this scheme becomes
incompatible with symmetry breaking after the two loop
contributions were included. In fact the freedom of
renormalization prescription choices will be further restricted
after imposing the stability condition for the solutions, which be
clear shortly in next subsection.
\subsection{Stability of symmetry breaking and the criterion for
coupling constant} From the above discussions, it is not clear yet
whether the symmetry breaking solutions are stable or not, i.e.,
we have not confronted our intermediately renormalized effective
potential with physical conditions or requirements, which
corresponds to solving the renormalized quantities in terms of
physical quantities. To this end, let us calculate the vacuum
energy density of the symmetry breaking phase. Using
Eq.~(\ref{SCH2}) we have
\begin{eqnarray}
\label{vac} E_{+}(\lambda,\mu)\equiv
V_{(2l)}(\sqrt{12\lambda\phi^2})|_{\phi^2=\phi_{+}^2}=
-\frac{(12\lambda\phi^2_{+})^2}{2(8\pi)^2}
[1+\frac{3\lambda}{2\pi^2}(3L_{+}-2)]
\end{eqnarray}with the symbols defined in the previous subsections.
Since the weak-coupling vacuum state ($\phi=0$) energy is zero,
for the symmetry breaking states to be stable, we must require
that
\begin{eqnarray}
E_{+}(\lambda,\mu) \leq 0,
\end{eqnarray} that is,
\begin{eqnarray}
\label{stablecondition} L_{+} \geq
\frac{2}{3}-\frac{2\pi^2}{9\lambda}.
\end{eqnarray}
This criterion turns out to be a requirement of the renormalized
coupling constant, i.e.,
\begin{eqnarray}
\label{critical} \lambda\geq \hat{\lambda}_{cr}\equiv
\frac{4\pi^2}{\sqrt{4-36\alpha-27}-1},\ (>\lambda_{cr}=
\frac{4\pi^2}{\sqrt{4-36\alpha}-1}).
\end{eqnarray}
In all the schemes with symmetry breaking, the two critical values
of the coupling constant are greater than 1, we can conclude that
symmetry breaking could not happen in weak-coupling regime. The
critical couplings in various prescriptions are exhibited in the
Table~\ref{table2}.

\begin{table}[t]
\caption{Critical values of coupling constant in various schemes}
\begin{center}
\begin{tabular}{c|c|c} \hline \hline &&\\[-.4cm]
 Scheme & $\lambda_{cr}$&$\hat{\lambda}_{cr}$\\ \hline &&\\[-.4cm]
$\overline{MS}$&  4.368& $5.2024$\\
$\mu^2_{\Lambda}$& 5.1152 & 6.5797 \\
Jackiw & 6.5797 & 10.698\\
\hline \hline
\end{tabular}
\end{center}
\label{table2}
\end{table}

Now we see that dynamical symmetry breaking does happen in certain
renormalization schemes in the strong-coupling regime. On the
other hand, the stability requirement also imposes further
constraint on the prescription choices in order to predict
symmetry breaking. In this connection, note that the stable
condition~(\ref{stablecondition}) amounts to the following
mathematical requirement
\begin{eqnarray}
(1+\frac{4\pi^2}{\lambda})^2\leq-23-36\alpha.
\end{eqnarray}Since the left-hand side of this inequality could
not be less than $1^{+}$, then we obtain the following criterion
for $\alpha$, or for scheme choices,
\begin{eqnarray}
\label{cr2}
\alpha\leq-\frac{2}{3},
\end{eqnarray}which is more stringent requirement than
$\alpha<\frac{1}{12}$.
\subsection{RG invariance of vacuum energy and beta function}
Since the vacuum energy is a physical entity, it must be
renormalization group invariant, i.e., insensitive to the choice
of subtraction point within a scheme,
\begin{eqnarray}
\label{RGI} \mu\frac{d E_{+}(\lambda,\mu)}{d\mu}=0.
\end{eqnarray}We must stress that this condition in fact
defines a fundamental physical scale as input in this broken phase
that should be obtained from some kind of experimental
measurements, corresponding to the important and necessary step
after renormalization is done, i.e., to confront the renormalized
amplitudes with experiments or other physical inputs or conditions
where the physical scales come from\cite{Sterman}. Consequently, a
fundamental physical scale is introduced into the effective
potential.

From this equation, we can determine the beta function of
$\lambda$ as was did in ref.\cite{Stevenson}. First, let us
rewrite the vacuum energy density as
\begin{eqnarray}
\label{vac2} E_{+}=-\frac{\mu^4}{2(8\pi)^2}\varepsilon_{+}
(\lambda)e^{2L_{+}(\lambda)},
\end{eqnarray}with $\varepsilon_{+}(\lambda)\equiv1+
\frac{3\lambda}{2\pi^2}(3L_{+}(\lambda)-2)=\frac{3\lambda}{4\pi^2}
(\sqrt{\Delta}-3)$. Then we find from Eq.~(\ref{RGI}) that
\begin{eqnarray}
4\varepsilon_{+}(\lambda)e^{2L_{+}(\lambda)}+
\{\varepsilon_{+}(\lambda)e^{2L_{+}(\lambda)}\}^{\prime}
\beta(\lambda)=0,
\end{eqnarray} or equivalently,
\begin{eqnarray}
\beta(\lambda)\equiv \mu\frac{d \lambda}{d\mu}
=-4\frac{\varepsilon_{+}(\lambda) e^{2L_{+}(\lambda)}}
{\{\varepsilon_{+}(\lambda)e^{2L_{+}(\lambda)}\}^{\prime}}.
\end{eqnarray}Since $\varepsilon_{+}(\lambda)$ is positive
definite provided the symmetry breaking solution is stable,
\begin{eqnarray}
\{\varepsilon_{+}(\lambda)e^{2L_{+}(\lambda)}\}^{\prime}&=&
\{\frac{\varepsilon_{+}(\lambda)}{\lambda}(1+
\frac{4\pi^2}{9\lambda})+ \frac{(1+4\pi^2/\lambda)}{3\lambda}\}
e^{2L_{+}(\lambda)}>0
\end{eqnarray} and hence the beta function is negative definite
as long as the broken phase is stable,
\begin{eqnarray}
\beta(\lambda)=-\frac{12\lambda \varepsilon_{+}(\lambda)}
{\{\varepsilon_{+}(\lambda)(3+\frac{4\pi^2}{3\lambda})+
1+4\pi^2/\lambda\}}<0.
\end{eqnarray}

This is true for all three schemes allowing for symmetry-breaking
solution. When the coupling becomes infinitely strong, i.e.,
$\lambda\rightarrow\infty$, the beta function approaches a
straight line:
\begin{eqnarray}
\beta(\lambda)|_{\lambda\rightarrow\infty}\longrightarrow
-4\lambda,
\end{eqnarray} while when the coupling approaches the critical
value $\hat{\lambda}_{cr}$, the beta function also approaches a
straight line with the same ratio:
\begin{eqnarray}
\beta(\lambda)|_{\lambda\rightarrow \hat{\lambda}^{+}_{cr}}\sim
-4(\lambda-\hat{\lambda}_{cr}),
\end{eqnarray}

\begin{figure}[t]
\begin{center}
\vspace*{0cm} \hspace*{0cm} \epsfxsize=10cm \epsfbox{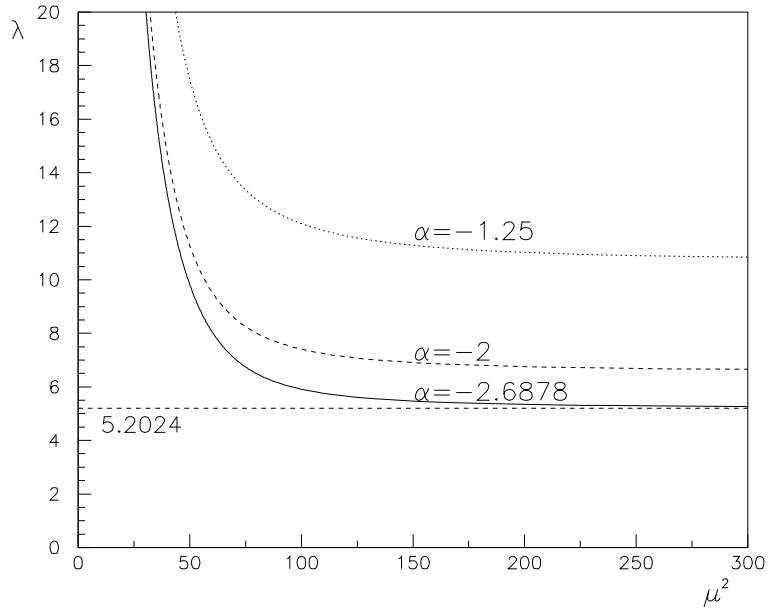}
\vspace*{-2.6cm} \caption{The running behavior of the coupling
constant in various prescriptions. The UV fixed points can be
found as the asymptotic lines. We have exhibited the UV fixed
point for the $\overline{MS}$ case.} \label{mu}
\end{center}
\end{figure}

The wonderful thing that enhances our faith in the two loop
effective potential is that all schemes (except the
Coleman-Weinberg scheme) predict the same kind of running behavior
of the coupling (the same kind of beta function)\footnote{This is
true in fact for all the prescriptions as long as the criterion
~(\ref{cr2} is satisfied, since the beta function is basically the
same except the UV fixed point varies with prescription.}, and we
could roughly imitate the true beta function with the following
qualitative approximation:
\begin{eqnarray}
\label{rude} \beta_{appr}(\lambda)=-4(\lambda-\hat{\lambda}_{cr}),
\ \ \ \lambda \in (\hat{\lambda}_{cr},\infty)
\end{eqnarray}with the obvious solution
\begin{eqnarray}
\label{rude2} \lambda-\hat{\lambda}_{cr}=\frac{\mu^4_0}{\mu^4}, \
\ \ \ \mu\in(0,\infty)
\end{eqnarray}which could also be obtained as a crude approximation
of Eq.~(\ref{vac2}). The RG-invariant scale $\mu^2_{0}$ should be
a function of the vacuum energy density as the fundamental
physical scale for this theory. Moreover, the running is
relatively milder in the UV region, which means that the coupling
constant does not become very large at energies that are not too
low. The true running behavior defined by Eq.~(\ref{vac2}) has
been plotted in Fig.~\ref{mu}.

Now it is clear that we obtained a \emph{nontrivial} theory with a
nonzero UV fixed point, $\hat{\lambda}_{cr}$, a strong coupling,
as is clear from Table~\ref{table2}, in contrast to the one loop
case. From Eq.~(\ref{rude2}) we can identify an IR pole in terms
of $\mu^2$, unlike the IR Landau pole in QCD, thus it is new at
least in a theoretical sense. No matter what kind of phenomenon it
defines, it is clear that within the two loop effective potential,
the dynamical symmetry-breaking phase is nontrivial without
asymptotic freedom, which means this phase is a totally
strong-coupling phase. Since this property is true in a number of
renormalization prescriptions that satisfy the criterion
~(\ref{cr2}), we feel that it is at least an interesting
phenomenon that deserves further examination. We emphasize that
our derivation here has not employed any unconventional or special
assumptions or approximations, all the techniques and arguments
are well known and well established. From now on we denote this
solution as SCRDSB for strong coupling regime dynamical symmetry
breaking.

\subsection{Further details about the SCRDSB}
Before speculating on this SCRDSB solution, let us examine the
scale dependence patterns of the main quantities of interests.

First let us look at the order parameter of the symmetry-breaking,
i.e., the square vacuum expectation value of the scalar field
$\phi^2_{+}$ or equivalently $\Omega^2_{+}$. Inverting the
dimensionless function of coupling, we can express the running of
the coupling in the following form by taking the vacuum energy
density as fundamental in Eq.~(\ref{vac2}), i.e.,
\begin{eqnarray}
\mu^2=8\pi e^{-L_{+}(\lambda)}
\sqrt{\frac{-2E_{+}}{\varepsilon_{+}(\lambda)}}.
\end{eqnarray} Combining this relation with the definition of
$L_{+}$, we find the dependence of $\Omega^2$ upon the running
scale,
\begin{eqnarray}
\label{Omega} \Omega^2_{+}(\lambda)=\mu^2e^{L_{+}(\lambda)}=
8\pi\sqrt{-2E_{+}/\varepsilon_{+}(\lambda)},
\end{eqnarray}or, equivalently,
\begin{eqnarray}
\phi^2_{+}(\lambda)=\frac{2\pi\sqrt{-2E_{+}/\varepsilon_{+}(\lambda)}}
{3\lambda}.
\end{eqnarray}Since the coupling runs, the order parameter also
runs from its dependence upon $\varepsilon_{+}(\lambda)$,
therefore we need to study the running behavior of
$\varepsilon_{+}(\lambda)$. Bearing in mind the running behavior
described in Eq.~(\ref{rude2}), we have,
\begin{eqnarray}
\varepsilon_{+}(\lambda)\|_{\lambda\rightarrow\infty}\rightarrow
\frac{3\lambda}{4\pi^2}\sqrt{(1-12\alpha)};\ \ \ \ \
\varepsilon_{+}(\lambda)\|_{\lambda\rightarrow
\hat{\lambda}^{+}_{cr}}\rightarrow \frac{\delta(\delta-1)}
{12\pi^2}(\lambda-\hat{\lambda}^{+}_{cr}),
\end{eqnarray}with $\delta\equiv \sqrt{4-36\alpha-27}$
being positive definite all the three schemes compatible with
symmetry breaking. Using Eq.~(\ref{rude2}) we find that in both IR
and UV regions,
\begin{eqnarray}
&&\varepsilon_{+}(\mu)\propto \frac{1}{\mu^4}.
\end{eqnarray}Then we obtain the asymptotic behaviors of the order
parameter $\Omega^2_{+}$ in both IR and UV regions
\begin{eqnarray}
\Omega^2_{+}(\mu)\propto \mu^2.
\end{eqnarray}But the asymptotic behaviors of $\phi^2_{+}$ is
somewhat different,
\begin{eqnarray}
\phi^2_{+}(\mu)\|_{\mu\rightarrow\infty}\propto \mu^2;\ \ \ \ \ \ \
\phi^2_{+}(\mu)\|_{\mu\rightarrow0}\propto \mu^6,
\end{eqnarray}which means that the square vacuum expectation
value of field vanishes more rapidly than $\Omega^2_{+}$. We note
that due to the extra term of $L_{+}$ in the vacuum energy
density, the parameter $\Omega^2$ is no longer a
RG-invariant\cite{Stevenson}, in contrast to the one loop
effective potential case.

\begin{figure}[t]
\begin{center}
\vspace*{0cm} \hspace*{0cm} \epsfxsize=9cm \epsfbox{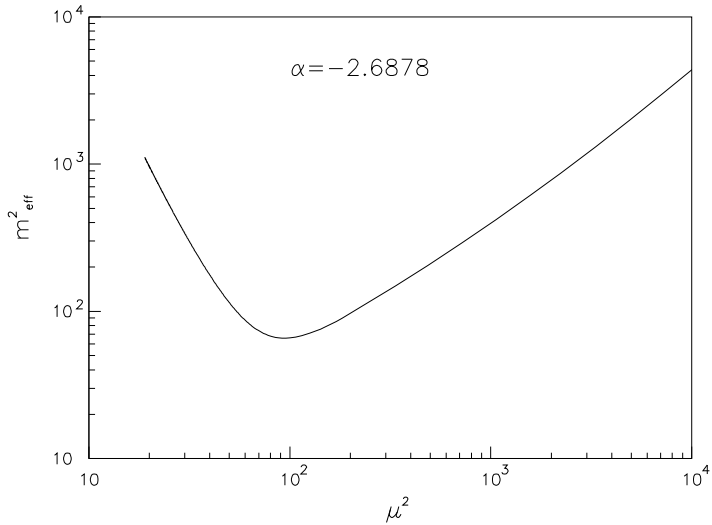}
\vspace*{-2.8cm} \caption{The running behavior of the effective
mass in $\overline{MS}$ scheme with $-2E_{+}$ set to 1.}
\label{meff}
\end{center}
\end{figure}

Similarly, we can obtain the asymptotic behavior of the effective
mass defined in Eq.~(\ref{mass}). Using Eq.~(\ref{Omega}), we have
\begin{eqnarray}
m^2_{eff}(\lambda)= \frac{18\lambda^2}{(2\pi)^2}
\sqrt{\frac{-2E_{+}}{\varepsilon_{+}(\lambda)}}.
\end{eqnarray}With the above preparations, we find that
\begin{eqnarray}
m^2_{eff}(\lambda)\|_{\lambda\sim\infty}\sim
\lambda^{\frac{3}{2}};\ \ \ \ \ \
m^2_{eff}(\lambda)\|_{\lambda\sim\hat{\lambda}_{cr}^{+}}
  \sim(\lambda-\hat{\lambda}_{cr})^{-\frac{1}{2}}.
\end{eqnarray} Or in terms of running scale,
\begin{eqnarray}
m^2_{eff}(\mu^2)\|_{\mu\sim0}\sim \frac{1}{\mu^6};\ \ \ \ \ \
m^2_{eff}(\mu^2)\|_{\mu\sim\infty}\sim\mu^2.
\end{eqnarray}
Here we found new asymptotic behaviors that differ from both the
asymptotic freedom and the triviality solutions. The effective
mass (self-energy at zero momentum) becomes singular at both IR
and UV ends. Only in the moderate energy region characterized by
the typical energy scale--the vacuum energy density--can we have
finite effective mass. (Of course we must be aware that since the
dynamics of SCRDSB exists entirely in a strong-coupling regime the
uncalculated higher-order loop corrections will probably change
the situation obtained here and make it even more complicated.)
The running behavior of the effective mass is plotted in
Fig.~\ref{meff}.

At this stage, one would naturally ask about the asymptotic
behaviors of the effective coupling, defined as $\lambda_{eff}
(\lambda)\equiv \frac{\partial^4 V_{(2l)}}{(\partial\phi)^4}
\|_{\phi^2=\phi^2_{+}}$. The dependence of this four-point vertex
function upon the renormalized coupling $\lambda$ reads
\begin{eqnarray}
\lambda_{eff}(\lambda)=\frac{3\lambda}{2\pi^4}\{16\pi^4+132
\pi^2\lambda+9(61+3\alpha)\lambda^2+9\lambda(4\pi^2+63\lambda)
L_{+}+81\lambda^2L^2_{+}\}.
\end{eqnarray}After some calculations, we have
\begin{eqnarray}
\lambda_{eff}(\lambda)\|_{\lambda\sim\infty}\sim10^1\lambda^3 ;
\ \ \ \ \ \lambda_{eff}(\lambda)\|_{\lambda\sim\hat{\lambda}_{cr}}
\sim 10^2 \hat{\lambda}_{cr},
\end{eqnarray} or equivalently
\begin{eqnarray}
\lambda_{eff}(\mu)\|_{\mu\sim0}\sim\frac{1}{\mu^{12}};
\ \ \ \ \ \lambda_{eff}(\mu)\|_{\mu\sim\infty}\sim10^3.
\end{eqnarray}Note that the effective coupling becomes more singular
than the effective mass does in the IR limit.

Since both the effective mass and the effective coupling become
extremely singular in the IR limit, it is not difficult to see
that in the low energy region, the kinetic energy of the scalar
field is negligible and the static potential energy dominates,
thus it seems impossible to find free scalar field quanta as
asymptotic states, in this sense the elementary scalar field seems
to be 'confined' somehow even in the high energy ranges. We might
detect some kind of bound states of such scalar field, with the
new bound states being also scalar states. So, even though we
found a scalar particle, there may appear another problem with
regard to whether the detected particles are elementary ones or
bound states of the elementary ones. In addition, the coupling is
still strong in the high energy region, though not infinitely
strong. The situation encountered here seems to indicate that the
Higgs model can allow for another scenario and symmetry-breaking
mechanism provided one explores it nonperturbatively. The Higgs
scalar quanta seem to be hidden 'heros' that did not like to be
'shown off' in the asymptotic states.

\section{A differential equation approach analysis}
Now we employ a new approach without explicit regulators or
deformations to calculate the loop diagrams. This approach is
based on the standard point of view that all the known QFT's are
effective theories for a completely well-defined quantum theory
containing 'correct' high-energy details\cite{PS2}. Then we should
make it clear that the UV structures of our present QFT's are
inevitably incorrect or inaccurate and should be replaced by the
'correct' underlying ones that are unknown to us yet and hence
certain diagrams can not be directly computed within the present
formulation of QFT's. (In conventional methods one introduces
artificial regularizations to imitate the underlying UV
structures.)

Fortunately, since differentiating a loop diagram with respect to
its 'low energy' parameters (momenta and mass(es) that
characterize the 'effective' QFT's) amounts to inserting 'low
energy' vertices to this diagram (this is valid in both the
underlying theory and the effective theories), which in turn
reduces the divergence degree of the diagram in terms of the
effective QFT's, we can compute a potentially divergent loop
diagram after differentiating them with respect to the (external)
momenta and/or mass(es) for appropriate times. In other words, we
can calculate the ill-defined diagrams by solving certain well
defined differential equations\cite{CGR}. In this approach, the
solutions would naturally contain unknown constants parametrizing
the ill-definedness or incompleteness (to be fixed by physical
'boundary conditions') of the effective theories. It is obvious
that this approach \emph{needs neither artificial regularizations
nor complicated procedures}.

\subsection{Recalculating the loop diagrams}
Now we demonstrate this method with the sunset diagram, the two
loop integral $I_{2}(\Omega)$.

(1). First, we differentiate it with respect to mass square
($\Omega^2$) for two times to remove all overall ill-definedness
(divergence),
\begin{eqnarray}
\label{DE}
\frac{\partial^2}{\partial_{(\Omega^2)^2}}I_{2}(\Omega)\equiv
6I_{2:(3;1;1)}(\Omega)+3I_{2:(2;2;1)}(\Omega)+
3I_{2:(2;1;2)}(\Omega)
\end{eqnarray}with
\begin{eqnarray}
I_{2:(\alpha;\beta;\gamma)}(\Omega)\equiv \int \frac{d^4k
d^4l}{(2\pi)^8(k^2+\Omega^2)^{\alpha}
((k+l)^2+\Omega^2)^{\beta}(l^2+\Omega^2)^{\gamma}}.
\end{eqnarray}The result is a sum of new diagrams without any
overall divergence. Among these diagrams, $
I_{\theta:(3;1;1)}(\Omega)$ still contains a subdivergence in the
$l$ integration\[ I_{(1;1)}(\Omega,k^2)\equiv \int
\frac{d^4l}{(2\pi)^4((k+l)^2+\Omega^2)(l^2+\Omega^2)}.\]

(2). Second, we treat this divergent subdiagram in the same way to
arrive at the following inhomogeneous differential equation
\begin{eqnarray}
\partial_{\Omega^2}I_{(1;1)}(\Omega,k^2)
=\frac{-1}{(4\pi)^2}\int^1_0 \frac{dx}{\Omega^2+(x-x^2)k^2}
\end{eqnarray}and its solution
\begin{eqnarray}
I_{(1;1)}(\Omega,k^2)=\frac{-1}{(4\pi)^2}\int^1_0 dx
\{\ln\frac{\Omega^2+(x-x^2)k^2}{\mu^2_{PDE}}+c_1\},
\end{eqnarray}with $c_1$ being the integration constants to be
fixed through physical 'boundary conditions'.

(3). Now we can compute the right hand side of Eq.~(\ref{DE}) and
obtain again an inhomogeneous differential equation as below
\begin{eqnarray}
\frac{\partial^2}{\partial_{(\Omega^2)^2}}I_{2}(\Omega)=
-\frac{3(\ln
\frac{\Omega^2}{\mu^2_{PDE}}+c_1-1)}{(4\pi)^4\Omega^2} ,
\end{eqnarray} and the solution to it reads
\begin{eqnarray}
I_{2}(\Omega)=-\frac{3}{2(4\pi)^4}\{\Omega^2[(\ln
\frac{\Omega^2}{\mu^2_{PDE}}+c_1-2)^2-(c_1-1)^2+1+
2c^{\theta}_1]+2c^{\theta}_2\}.
\end{eqnarray}with $\mu^2_{PDE}$,$c^{\theta}_{1}$ and
$c^{\theta}_{2}$being the constants (independent of masses,
coupling and momenta) to be fixed by 'boundary conditions'. The
single loop integrals can be done in the same way and are listed
in the Appendix A.

It is not difficult to see that, before fixing the constants, this
differential equation approach provides a most general
parametrization of the ill-defined loop diagrams. Any consistent
regularization and/or renormalization \emph{should be} a special
solution to these differential equations provided the
counter-terms are local functions of the momenta and mass(es). One
might feel that this approach is nothing but another form of the
powerful BPHZ program\cite{BPHZ}. To respond, we note the
following. First, one must employ a regularization method in BPHZ.
Second, the local terms in BPHZ are prefixed through the Taylor
expansion of the amplitudes, a crucial technical point, while in
the differential equation approach the local term are \emph{to be
fixed physically}. Third, the BPHZ ends up with the introduction
of infinite bare quantities while there is no room in principle
for such infinite quantities at all if one adopts the underlying
theory standpoint. Fourth, the application of BPHZ (and other
conventional programs) in nonperturbative circumstances is rather
involved that might preclude any useful (or trustworthy)
predictions\cite{NILOU}, while the differential equation approach
makes the calculation easier and the physical predictions more
accessible\cite{NILOU}.

In fact, one often relies on a good regularization method to make
the subtraction simpler, e.g., dimensional regularization for
gauge theories. Recently, it has been shown that in dimensional
regularization some subtraction is done implicitly without
introducing counterterms\cite{PH}. That is, we rely heavily upon a
regularization method that could discard divergences 'invisibly'.
If one disregards the underlying theory point of view where there
is no divergence but there are ambiguities, then there seems to be
no good reason to prefer the regularization methods that
\emph{simply discard some of the divergences without subtraction}.
For example, the modified minimal subtraction in dimensional
regularization does not lead to useful predictions in the
nonperturbative applications of the effective-field-theory
method\cite{EFT} to nucleon interactions\cite{Maryland}, which is
followed by the works that employ unconventional renormalization
methods\cite{Kaplan,Richardson,newren}. Applying the underlying
theory based differential equation approach will make the problem
easier to resolve\cite{055}.
\subsection{Relevance of the fixing
of the local ambiguities} Now we arrive at the following general
form of effective potential with unknown constants to be fixed,
\begin{eqnarray}
\label{VPDE} V^{(PDE)}_{(2l)}(\Omega )&=&\Omega
^{4}\{\frac{1}{144\lambda }+\frac{\hat{L}-\frac{3}{2}}{(8\pi
)^{2}}+\frac{3\lambda }{(4\pi )^{4}}[(\hat{L}-1)^{2}+
2(\hat{L}-2)^{2}-2(c_1-1)^{2}+2
+4c_{1}^{\theta }]\}\nonumber \\
&&+\Omega ^{2}\{\frac{c_{2}}{2(4\pi )^{2}}+\frac{6\lambda
(c_{2}[\hat{L}-1]+2c_{2}^{\theta })}{(4\pi )^{4}}\},
\end{eqnarray}where $\hat{L}=\ln
\frac{\Omega^2}{\mu^2_{PDE}}+c_1$ and all the $\phi$-independent
constant terms are discarded as they are irrelevant to our
discussions here. Naive dimensional analysis tells us that we have
three dimensional constants, $\mu^2_{PDE}$, $c_2$ and
$c^{\theta}_2$, and two dimensionless constants, $c_1$ and
$c^{\theta}_1$. In all the conventional prescriptions, the terms
quadratic in $\Omega$ are discarded somehow: In dimensional
regularization, it is done due to the vanishing (the 'invisible'
subtraction) of power divergences, while in cutoff regularization
it is just subtracted away by counterterms. Here we must fix it
via sound physical arguments.

We may expect that there should be at least a fundamental scale to
characterize the \emph{quantum} fluctuations of the scalar field.
As we are mainly concerned with the symmetry-breaking solution, we
temporarily take the vacuum energy density to play the role of the
fundamental scale. Generally speaking, all the three dimensional
constants should be of the same order of magnitude were they not
zero. Then the signs and magnitudes of $c_2$, $c^{\theta}_2$,
$c_1$ and $c^{\theta}_1$ will be crucial to the existence of
symmetry-breaking solutions. $c_1$ and $c^{\theta}_1$ can be put
into one constant $\alpha$ as this will not change the problem,
then
 Eq.~(\ref{VPDE}) becomes
\begin{eqnarray}
\label{VPDE2} V^{(PDE)}_{(2l)}(\Omega )&=&\Omega
^{4}\{\frac{1}{144\lambda }+\frac{\hat{L}-\frac{3}{2}}{(8\pi
)^{2}}+\frac{3\lambda }{(4\pi )^{4}}[(\hat{L}-1)^{2}+
2(\hat{L}-2)^{2}+\alpha]\}\nonumber \\
&&+\Omega ^{2}\{\frac{c_{2}}{2(4\pi )^{2}}+\frac{6\lambda
(c_{2}[\hat{L}-1]+2c_{2}^{\theta })}{(4\pi )^{4}}\}.
\end{eqnarray}With the presence of $c_2$ and $c^{\theta}_2$, we
will find that no matter what number we assign to $\alpha$, there
might be symmetry-breaking in this effective potential provided
the $c_2$ and $c^{\theta}_2$ are appropriately chosen, say,
$c_2>0, c^{\theta}_2=0$. This is because when $\Omega$ becomes
vanishingly small the potential reduces to
\begin{eqnarray}
V^{(PDE)}_{(2l)}(\Omega )\sim c_{2}\frac{6\lambda\Omega ^{2}
\hat{L}}{(4\pi )^{4}}
\end{eqnarray} where $\phi=0$ is a local maximum, a clear evidence
of symmetry breaking, which is true even if $c^{\theta}_2$ is not
zero as long as it is not too large compared to $c_2$. Of course,
if we let both $c_2$ and $c^{\theta}_2$ equal to zero, then
$\alpha$ will determine the existence of symmetry breaking
solutions.
\subsubsection{The rescaling insensitivity requirement and fine tuning}
The most important point is that if one adopts a fixing
prescription so that the quadratic terms are present, then we can
by no way remove them by redefinition of the coupling constant
(and perhaps $\phi$) without altering the symmetry breaking
status. That means, the fixing schemes with quadratic terms are at
least inequivalent to those without. As the underlying theory is
still unknown, we have to resort to experimental or other physical
means to fix them. Of course for such unrealistic model,
experimental data are unavailable, thus we need to search for
physical arguments. In the absence of obvious good clues to use, a
tentative argument might be that, due to the presence of the
dimensional constants $c_2$ and $c^{\theta}_2$ as the coefficient
of the quadratic terms, the effective potential would be rather
sensitive to the rescaling of these dimensional constants, in
contrast to the relatively milder rescaling behavior described by
the logarithmic dependence upon the dimensional constant
$\mu^2_{PDE}$. Then for the 'low energy' effective potential to be
less sensitive to the rescaling of the underlying details, we must
fix the dimensional constants $c_2$ and $c^{\theta}_2$ to be zero.

One might argue that this is just the unnatural fine tuning. If
the differential equation approach is taken as another way to
'renormalize' QFT's, this is true. However, if we adopt the
underlying theory point of view, we feel that this is a very
natural argument. This is because in the underlying-theory-based
differential equation there are no divergences to be subtracted
but only ambiguities to be fixed (a big improvement), then letting
these dimensional constants equal to zero based on the
insensitivity argument is just like what we usually do in solving
the Laplace equation or Schr\"odinger equation, namely imposing
sound boundary conditions. Thus the underlying theory and
differential equation approach offers a new way of understanding
the vanishing of the quadratic divergence: not from symmetry
argument but from the insensitivity of the effective theories'
quantities to the rescaling of the underlying structures
(represented by the arbitrary constants).
\subsubsection{Relevance of dimensionless constant(s)}
Of course there might be other possibilities. We will no longer
investigate this topic here. Now let us temporarily adopt the
rescaling insensitivity requirement and focus on the other
constants in the effective potential, i.e., $\mu^2_{PDE}$ and
$\alpha$ in the following form of the effective potential,
\begin{eqnarray}
\label{PDE3} V^{(PDE)}_{(2l)}(\Omega )&=&\Omega
^{4}\{\frac{1}{144\lambda }+\frac{\tilde{L}-1/2}{(8\pi
)^{2}}+\frac{3\lambda }{(4\pi )^{4}}[\tilde{L}^{2}+
2(\tilde{L}-1)^{2}+\alpha]\}
\end{eqnarray}where $\tilde{L}=\ln
\frac{\Omega^2}{\mu^2_{PDE}}+c_1 -1$. Now since $\alpha$ is
dimensionless and $\mu^2_{PDE}$ only appears in the logarithmic
functions, the rescaling insensitivity requirement is basically
satisfied (which is just the variant form of renormalization group
invariance). However, this requirement does not automatically
avoid the additional 'sensitivity' to the definition of the
dimensionless constant $\alpha$ (or $c_1$ and $c^{\theta}_1$), the
reason has already been given in section two.

\subsubsection{Non-existence of asymptotic freedom}
As a by product we can determine whether the UV fixed point could
be zero within the two loop effective potential. Here is the
reasoning. In order to get the asymptotic free solution, i.e.,
$\hat{\lambda}_{cr}=0$, it is clear from Eq.~(\ref{critical}) that
we must require the constant $\alpha$ to be infinitely large,
\begin{eqnarray}
\frac{4\pi^2}{\sqrt{4-36\alpha-27}-1}=0 \rightarrow
\alpha=-\infty.
\end{eqnarray} This is in fact a divergent constant. No sensible
renormalization prescription could allow for such a divergent
number. If one adopts the underlying theory point of view, it is
also an unacceptable choice of definition. Otherwise, it might
imply the underlying structures do not decouple with the 'low
energy' effective theories. Therefore we conclude that the UV
fixed point at two loop level can not be zero, i.e., the solution
can not be an asymptotic one, if we accept the parametrization
Eq.~(\ref{PDE3}) or Eq.~(\ref{SCH}). Generally the magnitude of
$\alpha$ should be of order not too bigger than $10^2$, then the
magnitude of the UV fixed point value of $\lambda$ should be
around $\frac{4\pi^2}{60}\sim 0.6$, that is, roughly of the order
1, which means that the broken phase can not be a weak coupling
one even in the high-energy region.

\section{Discussions and summary}
To recapitulate, in Secs. II and III, we made use of the
well-known two loop calculations to search for the
symmetry-breaking solutions. Our results here are new in two
aspects. (i) First, the striking prescription dependence of the
nonperturbative framework differs from that of the standard
perturbative framework\cite{scheme}, in other words, the
perturbative scheme dependence pattern is no longer valid in
nonperturbative contexts. Thus we can understand the relevance of
prescription found in nonperturbative applications like in
Ref.\cite{Kaplan,Maryland,Steele,Richardson} and \cite{NILOU};
(ii). Second, we found (in a number of renormalization
prescriptions) that the massless $\lambda\phi^4$ model could also
allow for a totally (non-perturbative) strong coupling dynamics
regime with negative beta function (SCRDSB), and therefore could
be nontrivial, at least in the two loop effective potential.

Although this phenomenon (SCRDSB) is only discovered in the two
loop effective potential, we found at least there is one thing
that is in common with the one loop case: the existence of
nontrivial phase of dynamics with broken symmetry that is strongly
coupled at least in the IR region. Considering the new kind of
diagrams beginning to appear from two loop level (the sunset
diagram and so on), such 'consensus' is conspicuous. We think
nontrivial solution might persist after including still
higher-order contributions, with the running behaviors being more
complicated, perhaps with more stringent constraints on the scheme
choices.

As far as the two loop effective potential is concerned, it is
very difficult to define asymptotic final states, thus the scalar
field theories with quartic interaction is rather different from
the gauge theories: it may have a broken phase that exists
entirely in the strong-coupling regime. Thus such scalar field
theories with quartic interactions might not permit the elementary
scalar fields to appear in the final asymptotic states. This
scenario might be of certain reference value to Higgs physics.

Another main task that has been performed is that we reanalyzed
the loop diagrams from the underlying theory point of view, which
takes all the presently known QFT's that suffer UV ill-definedness
to be ill-defined formulations of the effective 'low energy'
sectors. Then we showed clearly that the prescriptions or choices
for fixing the local ambiguities \emph{are} relevant to physical
properties encoded in the nonperturbative effective potential,
especially for the quadratic terms. In contrast to the
conventional regularization and renormalization programs where
power divergences are present and must be subtracted carefully
(fine tuning), in the underlying theory understanding, we can fix
them to be zero under the insensitivity requirement. This is a
natural procedure that is usually done in electrodynamics and
quantum mechanics, i.e., imposing appropriate boundary conditions
on the solutions obtained from the Laplace or Schr\"odinger
equation. In this way, we arrive at a new understanding of the
naturalness problem.

Furthermore, we also showed that there could not be reasonable
prescriptions that would allow for asymptotic freedom in the
broken phase as long as the two loop effective potential is
considered. In the underlying theory understanding, this is also
true.

Since more efforts in the realistic model are needed, we will
refrain here from making further comments. Our only aim here is to
draw attention to the reexamination of our triviality conviction
about the $\lambda\phi^4$ model and to the investigation of its
new nonperturbative properties (the perturbative regime is
unavoidably trivial).

In summary, we reconsidered the massless $\lambda\phi^4$ model
with $Z_2$ symmetry and found that at two loop level the
nonperturbative effective potential's predictability of symmetry
breaking depends upon the renormalization prescriptions in use.
The prescription used by Coleman and Weinberg in their pioneering
work\cite{CW} was shown to be incompatible with symmetry breaking
in the two loop effective potential, while the modified minimal
subtraction in dimensional regularization, Jackiw's prescription,
and others were shown to be able to accommodate symmetry-breaking
solution in the two loop effective potential. The reason for the
relevance of prescriptions in nonperturbative contexts was given.
The potential was also recalculated and reanalyzed in a
differential equation approach based on the standard point of view
that a complete theory underlies all the QFT's that suffer UV
divergences. The relevance of the prescriptions for fixing the
local ambiguities was stressed and the rationality of this
approach was highlighted.
\section*{Acknowledgement}
The authors are grateful to W. Zhu for critical discussions and
encouragements. We also benefited from conversations with Prof.
Xun Xue. This work is supported in part by the National Natural
Science Foundation of China under Grant No. 10075020.

\section*{Appendix A}
In this appendix, we write down all the needed one loop integrals
calculated respectively in dimensional, cut-off and differential
equation approach.
\begin{eqnarray}
\mu^{2\epsilon}I^{(D)}_{0}(\Omega)&=&\int\frac{\mu^{2\epsilon}d^D
k}{(2\pi)^D}\ln(1+\frac{\Omega^2}{k^2})=
-\frac{\Omega^4\Gamma(1+\epsilon)}
{(4\pi)^2\epsilon(1-\epsilon)(2-\epsilon)}
(\frac{4\pi\mu^2}{\Omega^2})^{\epsilon};\\
\mu^{2\epsilon}I^{(D)}_1(\Omega)&=&\int\frac{\mu^{2\epsilon}d^D
k}{(2\pi)^D}\frac{1}{k^2+\Omega^2}=
-\frac{\Omega^2\Gamma(1+\epsilon)}
{(4\pi)^2\epsilon(1-\epsilon)}(\frac{4\pi\mu^2}{\Omega^2})^{\epsilon};\\
I^{(\Lambda)}_{0}(\Omega)&=&\int_{\Lambda}\frac{d^4k}
{(2\pi)^4}\ln(1+\frac{\Omega^2}{k^2})\nonumber \\
&=&\frac{1}{2(4\pi)^2}\{\Lambda^4\ln
\frac{\Lambda^2+\Omega^2}{\Lambda^2}-\Omega^4
\ln\frac{\Lambda^2+\Omega^2}{\Omega^2}+\Lambda^2\Omega^2\};\\
I^{(\Lambda)}_{1}(\Omega) &=& \int_{\Lambda}\frac{d^4k}
{(2\pi)^4}\frac{1}{k^2+\Omega^2}=\frac{1}{(4\pi)^2}\{-\Omega^2\ln
\frac{\Lambda^2+\Omega^2}{\Omega^2}+\Lambda^2\} ;\\
I_{0}(\Omega)&=&\frac{1}{(4\pi)^2}\{\frac{\Omega^4}{2}[\ln
\frac{\Omega^2}{\mu^2_{PDE}}+c_1-3/2]+c_2\Omega^2+c_3\};\\
I_{1}(\Omega)&=&\frac{1}{(4\pi)^2}\{\Omega^2[\ln
\frac{\Omega^2}{\mu^2_{PDE}}+c_1-1]+c_2\}.
\end{eqnarray}
Note that in the differential equation approach there appear
unknown constants in the integrals parametrizing the
ill-definedness. The constants should be fixed by 'boundary
conditions' as discussed above.
\section*{Appendix B}
In this section, we describe the $\mu^2_{\Lambda}$ scheme that
mimics the $\overline{MS}$ scheme in dimensional regularization,
i.e., we merely subtract the cut-off containing parts. Let us
demonstrate it with the sunset diagram.

First let us list out all the relevant integrals or diagrams,
\begin{eqnarray}
I^{(\Lambda)}_{1}(\Omega) &=&\frac{1}{(4\pi)^2}\{-\Omega^2\ln
\frac{\Lambda^2+\Omega^2}{\Omega^2}+\Lambda^2\}=\frac{1}{(4\pi)^2}
\{ \Lambda^2-\Omega^2\ln
\frac{\Lambda^2}{\Omega^2}+o(\Lambda^{-2})\}; \\
I^{\Lambda}_{(1;1)}(\Omega,k^2) &=&\frac{1}{(4\pi)^2}(\int^1_0 dx
\ln \frac{\Lambda^2}{\Omega^2+x(1-x)k^2}-1)+o(\Lambda^{-2});\\
I^{(\Lambda)}_{2}(\Omega)&=&\frac{1}{(4\pi)^4}\{2\Lambda^2
-\frac{3\Omega^2}{2}\ln^2\frac{\Omega^2}{\Lambda^2}+3\Omega^2\ln
\frac{\Omega^2}{\Lambda^2}+o(\Lambda^{-2})\}.
\end{eqnarray}The counter term for sub-divergence in
$I^{(\Lambda)}_{2}(\Omega)$ comes from the log in
$I^{\Lambda}_{(1;1)}(\Omega,k^2)$ containing $\Lambda$ as an
argument, i.e., from $\ln \frac{\Lambda^2}{\mu^2}$ together with
factors from graph topology and angular integration. Thus the
counterterm for the sunset diagram reads
\begin{eqnarray}
c.t.(1)&=&\frac{(12\lambda)^2}{(4\pi)^2}\phi^2(\ln\frac{\Lambda^2}{\mu^2})
\times
I^{(\Lambda)}_1(\Omega)=\frac{12\lambda\Omega^2}{(4\pi)^4}\{\Omega^2\ln
\frac{\Lambda^2}{\mu^2}\ln\frac{\Omega^2}{\Lambda^2}+
\Lambda^2\ln\frac{\Lambda^2}{\mu^2}\}.
\end{eqnarray}
Here we selected an arbitrary scale to balance the dimension in
the argument of log. After removing the subdivergence, we get for
the sunset diagram
\begin{eqnarray}
-48\lambda^2\phi^2I^{(\Lambda)}_{2}(\Omega)+c.t.(1)&=&
\frac{6\lambda\Omega^4}{(4\pi)^4}\{(\ln
\frac{\Omega^2}{\Lambda^2}-1)^2-1\}-
\frac{8\lambda\Lambda^2\Omega^2}
{(4\pi)^4}\nonumber \\
&&+\frac{12\lambda\Omega^2}{(4\pi)^4}\{\Omega^2\ln
\frac{\Lambda^2}{\mu^2}\ln\frac{\Omega^2}{\Lambda^2}+
\Lambda^2\ln\frac{\Lambda^2}{\mu^2}\}\nonumber \\
&=&\frac{6\lambda\Omega^4}{(4\pi)^4}\{(\ln
\frac{\Omega^2}{\mu^2})^2-2\ln
\frac{\Omega^2}{\mu^2}\}+\frac{6\lambda\Omega^4\ln
\frac{\Lambda^2}{\mu^2}}{(4\pi)^4}\{2-(\ln
\frac{\Lambda^2}{\mu^2})\} \nonumber \\
&&+\frac{6\lambda\Lambda^2\Omega^2}{(4\pi)^4} \ln
\frac{\Lambda^2}{\mu^2}-\frac{8\lambda\Lambda^2\Omega^2}
{(4\pi)^4}.
\end{eqnarray}

Now we see all remaining divergences are purely 'local' and can be
removed through introducing a second counterterm $c.t.(2)$,
\begin{eqnarray}
c.t.(2)&=&-\frac{6\lambda\Omega^4\ln \frac{\Lambda^2}{\mu^2}}
{(4\pi)^4}\{2-(\ln \frac{\Lambda^2}{\mu^2})\}
-\frac{6\lambda\Lambda^2\Omega^2}{(4\pi)^4} \ln
\frac{\Lambda^2}{\mu^2}+\frac{8\lambda\Lambda^2\Omega^2}
{(4\pi)^4},
\end{eqnarray}which contains no finite part, and the renormalized
sunset diagram now takes the following form
\begin{eqnarray}
[-48\lambda^2\phi^2I^{(\Lambda)}_{2}(\Omega)]^{(\mu^2_{\Lambda})}
&\equiv&
-48\lambda^2\phi^2I^{(\Lambda)}_{2}(\Omega)+c.t.(1)+c.t.(2)
\nonumber \\ &=&\frac{6\lambda\Omega^4}{(4\pi)^4}\{(\ln
\frac{\Omega^2}{\mu^2})^2-2\ln \frac{\Omega^2}{\mu^2}\}.
\end{eqnarray}
\section*{Appendix C}
In this appendix, we verify that even in the original
parametrization the Coleman-Weinberg scheme is still incompatible
with the DSB solution. The two loop effective potential in Ref.
\cite{Jackiw} reads
\begin{eqnarray}
\tilde{V}^{(CW)}_{(2l)}&=&\frac{\lambda\phi^4}{4!}(1+a\lambda
l+a^2\lambda^2l^2+b\lambda^2l+a^2\lambda^2
\frac{205}{36});\nonumber \\
a&=&\frac{3}{32\pi^2};\ \ b=\frac{-3}{4(4\pi)^4};\ \
l=\ln\frac{\phi^2}{M^2}-\frac{25}{6}.
\end{eqnarray}
From the first order condition, we find the following equation
\begin{eqnarray}
2(1+a\lambda l+a^2\lambda^2l^2+b\lambda^2l+a^2\lambda^2
\frac{205}{36})+a\lambda+2a^2\lambda^2l+b\lambda^2=0,
\end{eqnarray} that is,
\begin{eqnarray}
2a^2\lambda^2l^2+(2a\lambda+2b\lambda^2+2a^2\lambda^2)l+2+a
\lambda+ b\lambda^2+a^2\lambda^2\frac{205}{18}=0.
\end{eqnarray}The corresponding delta reads
\begin{eqnarray}
\tilde{\Delta}&\equiv&(2a\lambda+2b\lambda^2+2a^2\lambda^2)^2-
8a^2\lambda^2
(2+a\lambda+b\lambda^2+a^2\lambda^2\frac{205}{18})\nonumber \\
&=&-(\frac{3\lambda}{16\pi^2})^2\{3+\frac{\lambda}{16\pi^2}
+\frac{195\lambda^2}{4(4\pi)^4}\}<0.
\end{eqnarray}This inequality implies the incompatibility of the
Coleman-Weinberg scheme with symmetry breaking at two loop level.

\end{document}